\documentclass[11pt, a4paper]{article}
\usepackage{jcappub}

\usepackage[utf8]{inputenc} 
\usepackage[T1]{fontenc}    
\usepackage{lmodern}
\usepackage{hyperref}       
\usepackage{url}            
\usepackage{amsfonts}       
\usepackage{amsmath,amsthm,amssymb} 
\usepackage{bm}
\usepackage{bbm}
\usepackage{dcolumn}
\usepackage{nicefrac}       
\usepackage{graphicx}
\usepackage{xcolor, etoolbox}
\usepackage{mathtools}
\usepackage{mathrsfs}
\usepackage{soul}
\usepackage{ragged2e}
\usepackage{microtype}      
\usepackage{braket}      
\usepackage{enumitem}
\usepackage{xspace}
\usepackage{dsfont}
\usepackage{float}
\usepackage[export]{adjustbox}

\DeclarePairedDelimiter\abs{\lvert}{\rvert}

\newcommand{\dd}{{\text{d}}}

\newcommand{\mpl}{\text{M}_{\text{P}}}

\newcommand{\ti}{t_{\text{i}}}
\newcommand{\tf}{t_{\text{f}}}
\newcommand{\etai}{\eta_{\text{i}}}
\newcommand{\etaf}{\eta_{\text{f}}}
\newcommand{\af}{a_{\text{f}}}

\newcommand{\dS}{\text{dS}}
\newcommand{\av}{\text{av}}
\newcommand{\td}{\text{today}}

\hypersetup{
    colorlinks,
    linkcolor=blue,
    citecolor=blue,
    urlcolor=blue,
    breaklinks=true 
}

\graphicspath{{./figures/}} 

\begin{document}

\title{\boldmath Estimation of gravitational production uncertainties}

\author[a]{Jose A. R. Cembranos,}
\author[a]{Luis J. Garay,}
\author[a,b]{Álvaro Parra-López}
\author[a, c]{and Javier Ortega del Río}

\affiliation[a]{Departamento de F\'isica Te\'orica and IPARCOS, Facultad de Ciencias F\'isicas, \\
Universidad Complutense de Madrid, Ciudad Universitaria, 28040 Madrid, Spain} 
\affiliation[b]{Instituto de F\'isica Te\'orica UAM/CSIC, calle Nicol\'as Cabrera 13-15,\\ Cantoblanco, 28049, Madrid, Spain}
\affiliation[c]{Department of Physics and EHU Quantum Center, University of the Basque Country UPV/EHU, 48080 Bilbao, Spain}

\emailAdd{cembra@fis.ucm.es}
\emailAdd{luisj.garay@ucm.es}
\emailAdd{alvaparr@ucm.es}
\emailAdd{javier.ortega@ehu.eus}
	
\date{\today}


\abstract{Gravitational production of scalar, non-minimally coupled dark matter depends on the specifics of the inflationary model under consideration. We analyze both Starobinsky inflation and a quadratic potential, solve the full background dynamics, study pair production during inflation and reheating, and find that the observed dark matter abundance can be explained solely by this mechanism, regardless of the inflationary model. Qualitative differences between the two cases only appear for dark matter masses close to the inflationary scale. In addition, we identify a large region in parameter space in which gravitational production of dark matter is mostly independent of the chosen inflationary potential, highlighting the robustness of this dark matter production mechanism and its independence of the unknown particular details of inflation. In the region of masses lower than the scale of inflation, and sufficiently away from the conformal limit, the total comoving number density of produced particles becomes a function of the coupling to the geometry alone. This allows us to provide an approximated analytic expression for fitting the resulting abundance.}

\keywords{Dark matter theory, Cosmology of Theories beyond the SM, Inflation, Quantum fields in curved spacetimes}

\arxivnumber{2507.02410}
~\hfill IPARCOS-UCM-25-062\par
~\hfill IFT-UAM/CSIC-25-73\par
\maketitle
\flushbottom

\section{Introduction}

The standard cosmological $\mathrm{\Lambda CDM}$ model includes dark matter as one of the components of the universe to explain a wide array of observations. However, its fundamental nature remains unknown. Due to the apparently negligible interaction between dark matter and the Standard Model components \cite{Gaskins_2016}, the mechanism by which it emerges remains an open question \cite{BAER20151,Ford2021}. In this context, gravitational production due to the expansion of the geometry \cite{Parker1968,Parker1969,Parker1971,Ford1987} in the early universe, i.e., during inflation \cite{Starobinsky1980, Guth1981, Linde1982} and reheating \cite{Kolb1990,Kofman1994,Kofman1997,Allahverdi2010,Baumann2015}, has been proposed as a mechanism for dark matter generation 
(see \cite{Ford2021,Kolb2021} for recent reviews on this subject). This phenomenon, rooted in quantum field theory in curved spacetime \cite{Birrell1982,Mukhanov2007,Parker2009}, affects all fields non-conformally coupled to the geometry, regardless of its interactions with other fields.

Cosmological inflation is a theoretical framework developed to address several issues inherent in the Standard Cosmological Model. Originally proposed to resolve the horizon and flatness problems \cite{Guth1981}, inflation involves a brief period of accelerated expansion, typically driven by the dynamics of a scalar field known as the inflaton. Quantum fluctuations of the inflaton field are stretched to cosmological scales during inflation and provide a natural origin for the primordial density perturbations. These perturbations seed the anisotropies observed in the cosmic microwave background (CMB) and underpin the large-scale structure of the Universe. For a variety of inflationary models, the predicted spectrum and statistical properties of these fluctuations show good agreement with current observational data. Additionally, inflation offers a compelling explanation for the apparent absence of magnetic monopoles by diluting their density to undetectable levels \cite{Guth1981}.

Following the inflationary epoch, the Universe undergoes a phase known as reheating, during which the energy stored in the inflaton field is transferred to Standard Model particles. This marks the beginning of the radiation-dominated era. During inflation and reheating, when the geometry suffers rapid changes, one expects a large production of this kind \cite{Cembranos2023,Alvarez2025}, which could constitute the totality of the observed dark matter abundance. Particle production remains significant throughout the reheating phase until the Universe transitions to an adiabatic regime, when gravitational production becomes negligible.

Original studies on gravitational dark matter production focused on super-heavy scalar candidates minimally coupled to gravity \cite{Chung1998a,Chung1998b,Kolb1998,Kuzmin1999,Chung2001,Chung2003}, which were later extended to the cases of non-minimal coupling \cite{Markkanen2018,Fairbairn2019,Tenkanen2019,Chung2019,Hashiba2019,Cembranos2020,Herring2020,Kainulainen2023,Garcia2023a,Garcia2023b,Cembranos2023,BorrajoGutierrez:2020zkp}, vector \cite{Graham2015,Ema2019,Bastero2019,Ahmed2020,Sato2022,Cembranos2024,Ozsoy2024,Capanelli2024a,Capanelli2024b,Capanelli2024c}, or fermionic fields \cite{Ema2019}. Recently, it has been shown that the tachyonic regimes induced by the oscillations of the curvature scalar have the ability to increase pair production by orders of magnitude \cite{Ema2016,Markkanen2017,Ema2018,Cembranos2023,Cembranos2024}. On the other hand, the weakly interacting nature of dark matter naturally leads to the assumption that, when gravitational production becomes irrelevant, the number density of dark matter particles produced in the early Universe is not modified by any process other than the dilution due to the expansion of spacetime. This allows to directly compare the predictions of gravitational dark matter production to the experimentally observed abundance, provided other cosmological constraints on, e.g., isocurvature perturbations or structure formation \cite{Garcia2024b}, are fulfilled.

Nevertheless, the specific properties of the dark matter candidates produced gravitationally (their mass and the characteristics of their coupling to the geometry) are highly dependent on the particular inflation model and the details of reheating, leading to significant uncertainties in predicting their abundance. To address this, in this work, we perform an estimation of the model dependence of gravitational particle production as a dark matter generation mechanism, focusing on the case of a scalar spectator field $\varphi$ non-minimally coupled to gravity in single-field, chaotic inflationary backgrounds.

The field $\varphi$ is assumed to be massive, initially in its vacuum state, and coupled to gravity through a term $\xi R \varphi^2$, where $R$ is the Ricci scalar. Its energy density remains subdominant throughout. This means that its backreaction on the background can be safely ignored. Particle production is computed non-perturbatively via the Bogoliubov formalism. To evaluate the impact of the inflationary model, we consider two representative inflaton potentials: the quadratic potential---useful for comparison with existing literature despite moderate observational tension \cite{Markkanen2017,Markkanen2018,Fairbairn2019,Kainulainen2023,Cembranos2023}---and the Starobinsky potential, which offers and excellent fit to CMB observations \cite{Starobinsky:1980te,lust2024starobinsky}. For both, we assume slow-roll dynamics and numerically solve the background equations to determine the evolution of the scale factor and curvature scalar, which drive the mode equation for the dark matter field. 

We carefully analyze in which regions of parameter space the choice of inflaton potential significantly affects the produced dark matter density, and study the dependence of the resulting abundance on the reheating temperature. Additionally, we derive an analytic approximation for the dark matter abundance valid in the regime of low mass and large curvature coupling, which incorporates the inflaton mass, and thus unifies both models under a common framework. With this, we aim at obtaining a generic estimation of the dark matter abundance originated by this mechanism across a class of single-field inflationary models.

To compute particle production, we solve the mode equation for the scalar field $\varphi$ taking advantage of the analytical approximation for the slow-roll regime derived in detail in \cite{Cembranos2023}, and, once the approximation is no longer valid, we continue evolving the modes numerically until particle production becomes negligible, i.e., when the expansion of spacetime becomes adiabatic. We will assume that the field is initially in the Bunch-Davies vacuum \cite{Bunch1978}, which is the natural vacuum for such a field at the onset of inflation, where the geometry approaches de Sitter. For the vacuum of an observer living after pair production has become inefficient, we will consider the customary adiabatic prescription \cite{Birrell1982}.  

The \textit{remainder} of this paper is as follows. In section \ref{sec:ScalarField} we describe the dynamics of a spectator scalar field non-minimally coupled to gravity in an FLRW background, focusing on its mode description, and introduce the formalism of Bogoliubov transformations which will allow us to compute pair production. Then, in section \ref{sec:backgr} we introduce both the quadratic and the Starobinsky inflationary models that determine the particular evolution of the geometry during the early Univese. In section \ref{sec:production} we discuss the dynamics of the field modes, and how to solve its equation of motion for computing the number density of produced particles. The corresponding results are shown in section \ref{sec:abundance} in the form of production spectra and abundance after diluting the density of pairs until today. Lastly, we elaborate our conclusions in section \ref{sec:conclusions}

\textit{Notation}. We set $\mpl = 1/\sqrt{G},\, \hbar=c=k_B=1$, and use  the signature $(-,+,+,+)$. We will use the quadratic inflation potential mass, fixed to $m_{\phi}=1.21\times 10^{13}\ \mathrm{GeV}$, as a reference energy scale for comparison between both models.

\section{Scalar quantum field in cosmological spacetime}
\label{sec:ScalarField}

In this section we will briefly describe the dynamics of a scalar field in an FLRW spacetime, as well as the canonical quantization procedure.

\subsection{Dynamics of a scalar field in FLRW geometry}
We will consider a massive scalar field $\varphi$ with non-minimal coupling to gravity through the Ricci curvature scalar $R$. Its corresponding action reads
\begin{equation}
    S=-\frac{1}{2}\int \dd^4 x\sqrt{-g}(\partial_{\mu}\varphi\partial^{\mu}\varphi+m^2\varphi^2+\xi R\varphi^2),
\end{equation}
where $g$ is the determinant of the metric, $m$ is the field mass, and $\xi$ denotes the strength of the curvature coupling. For cosmological cases like the one at hand, the geometry is well described by the Friedmann-Lemaître-Robertson-Walker (FLRW) metric. For simplicity, and in agreement with observations \cite{planck18}, we consider flat spatial sections. The metric is therefore  
\begin{equation}
    \dd s^2=a^2(\eta)(-\dd\eta^2+\dd\bm{x}^2),
\end{equation}
in the so-called conformal time $\eta$, related to cosmological time $t$ by the relation $a(\eta)\dd \eta \equiv  \dd t$.

The corresponding equation of motion reads \cite{Cembranos2023} 
\begin{equation} 
\ddot{\varphi} + D\frac{\dot{a}}{a}\dot{\varphi}- \left[\frac{\Delta}{a^2} + m^2 + \xi R\right]\varphi = 0
\label{eq:KGEoM}
\end{equation}
where the dot denotes derivative with respect to cosmological time $t$, $m$ is the mass of the field, $\xi$ is the coupling constant, and the scalar curvature is related to the scale factor by $R=6(\ddot{a}/a+\dot{a}/a^2)$. We will restrict the parameter $\xi$ to $1/6\leq \xi\leq 1$, since smaller $\xi$ can give rise to tachyonic instabilities in the initial state before inflation, and larger $\xi$ give qualitatively similar results at the expense of greater computational difficulties \cite{Cembranos2023}, until sufficiently large values, where backreaction begins to play a role. 

For convenience, we will work with the rescaled field $\chi = a\varphi$, which can be written in momentum space as the linear combination of modes $v_k(\eta)$
\begin{equation}
    \chi(\eta,\bm{x})=\int \frac{\dd^3\bm{k}}{(2\pi)^{3/2}}[a_{\bm{k}}v_k(\eta)+a^*_{\bm{-k}}v_k^*(\eta)]e^{- i\bm{k}\bm{x}}.
\label{eq:FieldExpansion}
\end{equation}
When introducing the above mode expansion into the equation of motion \eqref{eq:KGEoM}, one obtains the following equation for the mode functions $v_k$:
\begin{equation}
v_k^{\prime\prime}(\eta) + \omega_k^2(\eta)v_k(\eta)=0,
\label{eq:ModeEq}
\end{equation}
where 
\begin{equation}
\omega^2_k(\eta)=k^2+a^2(\eta)\left[m^2+\left(\xi-1/6\right)R(\eta)\right],
\label{eq:ModeFreq}
\end{equation}
and the prime denotes derivative with respect to conformal time. This equation is that of an oscillator with a time-dependent frequency. Due to the isotropy of the background, the frequency does not depend on the direction of the momentum $\bm{k}$, but only on its modulus $k=\sqrt{\bm{k}^2}$, and so do the corresponding solutions. Note that, for the particular case in which $\xi=1/6$, the Ricci scalar term drops. This is called conformal coupling, since the effects of the expansion vanish as $m \to 0$, in which limit pair production does not occur.

The mode equation \eqref{eq:ModeEq} has a two-dimensional space of complex solutions, and therefore the set \{$v_k,v^*_k\}$---for a given value of $k$---forms a basis of the space of solutions. An orthonormal basis compatible with the Klein-Gordon product associated with the equation of motion~\eqref{eq:KGEoM}~\cite{Birrell1982} fulfills the following condition on its Wronskian, 
\begin{equation}
    \text{Wr}[v_k,v_k^*] = v_k v_k^{*'}-v_k'v_k^*=i.
\label{eq:Wronskian}
\end{equation}
Any other normalized solution $u_k$ can be written as a the linear combination
\begin{equation}
    u_k = \alpha_k v_k + \beta_k v_k^*,
\end{equation}
where $\alpha_k$ and $\beta_k$ are the so-called Bogoliubov coefficients. Compatibility with \eqref{eq:Wronskian} requires that
\begin{equation}
    \abs{\alpha_k}^2 - \abs{\beta_k}^2 = 1.
\end{equation}
It is important to note that the choice of basis in the field expansion \eqref{eq:FieldExpansion} determines the set of coefficients that will be eventually promoted to creation and annihilation operators, as we will discuss in the following subsection.

\subsection{Quantization}

Canonical quantization follows by promoting the coefficients $a_{\bm{k}}$ and $a^*_{\bm{k}}$ to operators $\hat{a}_{\bm{k}}$ and $\hat{a}^{\dag}_{\bm{k}}$, with the usual commutation relations \cite{Mukhanov2007}
\begin{equation}
    [a_{\bm{k}},a^\dag_{\bm{k'}}]=\delta^{(3)}(\bm{k}-\bm{k'}), \qquad [a_{\bm{k}},a_{\bm{k'}}]=0, \qquad [a^{\dagger}_{\bm{k}}, a^{\dagger}_{\bm{k'}}]=0.  
\label{eq:CommRels}
\end{equation}

Thus, for a given basis of solutions $\{v_k,v_k^*\}$, one can find creation and annihilation operators $a_{\bm{k}}$, $a_{\bm{k}}^{\dag}$ and define a vacuum state $\ket{0}$ such that $a_{\bm{k}}\ket{0}=0$ and particle states in the usual way. However, a different basis $\{u_k,u_k^*\}$ will give another set of creation and annihilation operators $b_{\bm{k}}$, $b_{\bm{k}}^\dag$---and thus to another definition for the vacuum and particle states. These new $b$-operators are related to the $a$-operators by the same Bogoliubov coefficients that relate the different set of modes, through
\begin{equation} \label{bogo2}
    b_{\bm{k}}=\alpha_k a_{\bm{k}}+\beta_k^* a^\dag_{\bm{-k}}.
\end{equation}
The number density of $b$-particles in the $a$-vacuum is given by \cite{Mukhanov2007}
\begin{equation}
    \mathcal{V}^{-1}\bra{0^a}\hat{b}_{\bm{k}}^{\dagger}\hat{b}_{\bm{k}}\ket{0^a}=|\beta_k|^2/(2\pi)^3,
\label{eq:NumberDensity}
\end{equation}
where $\mathcal{V}$ is related to the volume of the geometry's spatial sections. The total density is obtained as the integral over all modes
\begin{equation}
    n=\int \frac{\dd^3\bm{k}}{(2\pi)^3}|\beta_k|^2=\int \frac{\dd k}{2\pi^2}k^2|\beta_k|^2,
\label{eq:TotalDensity}
\end{equation}
where the $\beta_k$ Bogoliubov coefficient can be computed as $\beta_k=i\text{Wr}[u_k,v_k]$. 

One could, in principle, impose certain criteria for selecting one particular Fock quantization. For example, in Minkowski spacetime, Poincaré invariance alone selects a preferred quantum vacuum: the Minkowski vacuum. However, in general, spacetime symmetries are not enough to select a unique quantization.

Let us now associate the $a$-particles to the notion of vacuum of an observer living before certain expansion of the geometry, and the $b$-particles to an observer after the expansion. Before and after the expansion, both observers will have preferred---and in general different---vacuum notions, since spacetime becomes static in these regions. If the field is initially in the vacuum state (as measured by the $a$-observer), then Eq. \eqref{eq:TotalDensity} corresponds precisely to the number density of pairs produced due to the spacetime dynamics, that the $b$-observer will measure after the expansion. This is precisely the quantity we are interested in computing. 

However, such static regimes are absent in Cosmology. Nevertheless, it is possible to find regions in which symmetries are approximately recovered, so that observers living there can define suitable vacuum notions, as we will see below. In the following section, we will describe the specific evolution of spacetime we will consider.

\section{Background dynamics}
\label{sec:backgr}

Let us now discuss the inflationary models that characterize the background evolution leading to the dark matter production. We consider a chaotic inflationary model sourced by a scalar inflaton field $\phi$, whose dynamics is characterized by the well-known equation 
\begin{equation}
    \ddot{\phi}+3H\dot{\phi}+\frac{\dd V}{\dd\phi}=0,
\label{eq:InfEq}
\end{equation}
where $V$ is the inflationary potential, and $H=\dot{a}/a$ is the Hubble parameter, determined by the Friedmann equation \cite{1922ZPhy...10..377F,num}
\begin{equation}
    H^2=\left(\frac{\dot{a}}{a}\right)^2=\frac{8\pi}{3\mpl^2}\rho=\frac{8\pi}{3\mpl^2}\left(\frac{\dot{\phi}^2}{2}+V(\phi)\right).
\label{eq:HubbleRate}
\end{equation}
We will consider two different potentials: A quadratic potential and Starobinsky's. The first is given by
\begin{equation}
    V_{\text{quad}}=\frac{1}{2}m^2_{\phi}\phi^2,
\end{equation}
where $m_{\phi}$ is the inflaton mass, which we use to set a characteristic energy scale throughout this work. The Starobinsky potential is described by \cite{Maeda:1987xf,lust2024starobinsky}
\begin{equation} 
    V_{\text{Star}}=L^4\left[1-\exp\left(-4\sqrt{\frac{\pi}{3}}\frac{\phi}{\mpl}\right)\right]^2,
\label{eq:StarobinskyPotential}
\end{equation}
with $L$ a free parameter. Note that at late times, for $\phi\ll \mpl$, the potential \eqref{eq:StarobinskyPotential} behaves as
\begin{equation}
V(\phi) \simeq \frac{1}{2}\bar{m}_{\phi}^2\phi^2,
\end{equation}
where we defined the inflaton mass for the Starobinsky model $\bar{m}_{\phi}$ as
\begin{equation}
    \bar{m}_{\phi}^2 = \frac{32\pi}{3}\frac{L^4}{\mpl^2} \simeq 2.37m_{\phi} = 2.87 \times 10^{13} \, \text{GeV}.
\end{equation}
Lastly, the curvature scalar $R$, which is needed for characterizing the mode frequency~\eqref{eq:ModeFreq}, can be written in terms of the inflaton field $\phi$ as
\begin{equation}
    R=\frac{8\pi}{\mpl^2}\left[4V(\phi)-\dot{\phi}^2\right].
\label{eq:Ricci}
\end{equation}

In order to obtain the evolution of the modes through the whole inflationary and reheating regimes, we need to numerically solve the inflaton equation \eqref{eq:InfEq} from some initial time $\ti$ until a final time $\tf$ sufficiently deep into the reheating phase following inflation---such that spacetime expands very slowly and production becomes negligible, as we will discuss below. For both potentials inflation is assumed to start in the slow-roll regime ($|V|\gg \dot{\phi}^2$) and thus the corresponding spacetime is quasi-de Sitter at the beginning. In this regime the inflaton decreases slowly until it reaches a minimum of the potential and starts oscillating. 

\begin{figure}[t!]
    \centering
    \includegraphics[scale=0.5]{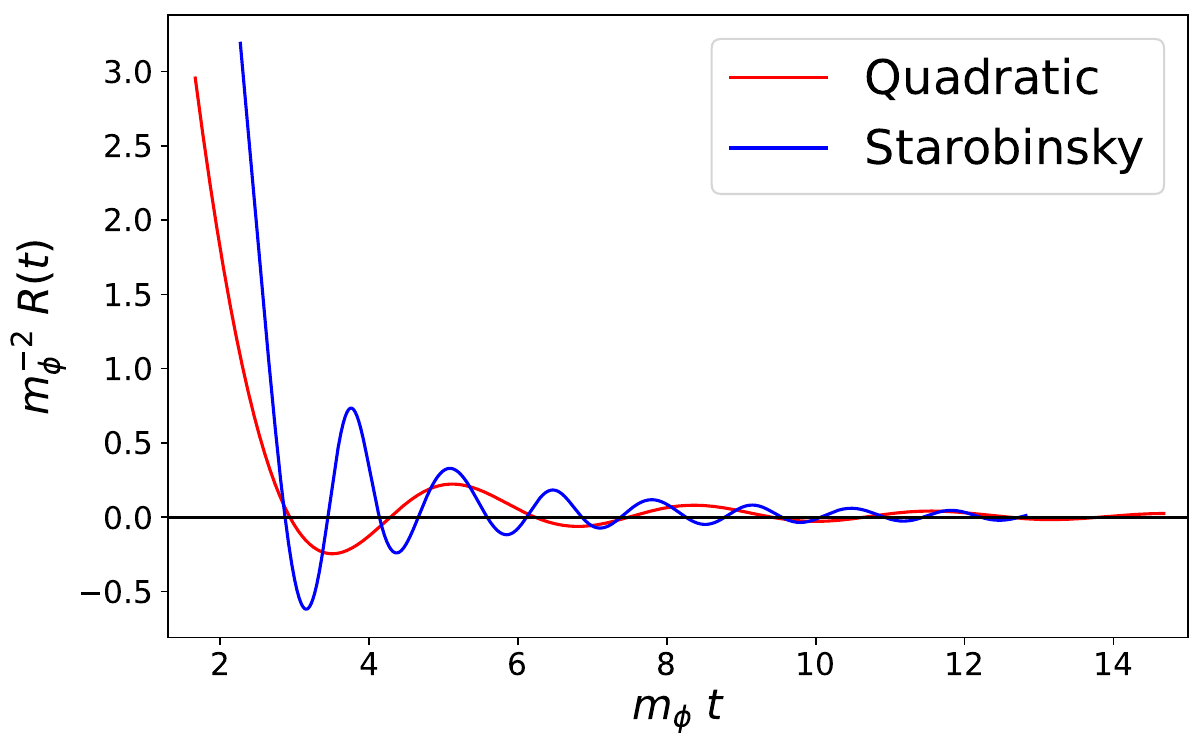}
    \caption{Ricci scalar $R$ as function of cosmological time $t$ during transition from inflation to reheating for the quadratic (red) and Starobinsky potential (blue).} 
    \label{fig:RicciScalar}
\end{figure}

Once we have a numerical solution for $\phi(t)$ and $\dot{\phi}(t)$, we integrate equation \eqref{eq:HubbleRate} to obtain the scale factor $a(t)$, up to an irrelevant normalization factor. Finally, we obtain an expression for the conformal time $\eta=\int \dd t/a(t)$ in terms of cosmological time $t$. This allows us to numerically find both $a(\eta)$ and $R(\eta)$, which encode all the background dependence of the mode frequency \eqref{eq:ModeEq}. The result is a quasi-exponential scale factor $a(\eta)$ for most part of the inflationary period until inflation is exited, when it approaches a matter-dominated form. Similarly, the Ricci scalar $R(\eta)$ remains almost constant for most part of inflation, falls and starts oscillating when reheating starts. This behavior is very similar for both the quadratic and Starobinsky potentials. Figure \ref{fig:RicciScalar} shows the time-evolution of the Ricci scalar close to the transition between inflation and reheating, for both inflationary potentials. We observe that the frequency of the oscillations in the Starobinsky case is larger, but otherwise the qualitative behavior is similar.

For solving the background, we choose the following initial conditions. For the quadratic potential, we solve the inflaton equation starting from $\ti = -15.35/m_{\phi}$, with $m_{\phi} = 1.21 \times 10^{13} \, \text{GeV}$. The initial conditions for the inflaton field are $\phi_{\text{i}} = 3\times 10^6 m_{\phi}$ and $\phi_{\text{i}}^{\prime} = 0$. For the Starobinsky potential, we take $t_{\text{i}} = -120/m_{\phi}$ and $\phi_{\text{i}} = 1.1 \times 10^6 m_{\phi}$, $\phi_{\text{i}}^{\prime} = 0$, with $L = 6.43 \times 10^{2} m_{\phi}$. The scale factor is normalized to $0.01$ at $\eta=0$ in the quadratic case, whereas in Starobinsky it is normalized to $175$ at the time $\etaf$ when we compute pair production, whose definition will be given below. The particular values of the time origin, as well as the normalization of the scale factors in each case are chosen for computational reasons and they do not change the physics. All values are chosen such that the inflationary models are compatible with observations \cite{planck18,Planckspectr}. 

\section{Particle production}
\label{sec:production}

Our goal is to compute the number density of produced particles due to the background dynamics described in the previous section, assuming that the system is initially in the vacuum state. This amounts to obtaining the Bogoliubov coefficient $\beta_k$ relating the notions of vacuum associated with observers living before and after the expansion, that is, at $\etai$ and $\etaf$, and computing \eqref{eq:TotalDensity}.

As mentioned before, while spacetime is expanding, one cannot, in general, define a preferred notion of vacuum. However, note that the geometry approaches de Sitter at the onset of inflation. De Sitter spacetime has enough symmetries such that one can select a preferred vacuum, the so-called Bunch-Davies vacuum \cite{Bunch1978}, which will be natural for an observer living at $\etai$\footnote{This is true in the case of a massive scalar field \cite{Mottola1986}}. Similarly, after entering reheating, spacetime dynamics slows down, and the expansion becomes more and more adiabatic. Eventually, the geometry becomes approximately static (when compared to the previous stages), and pair production becomes negligible. We define $\etaf$ as the time when this situation is reached. In this way, the number density of produced particles can be computed in an \textit{in-out} fashion.

For simplicity, we compute $\beta_k$ at $\etaf$, since its value is independent of the evaluation time. For this, we evolve the \textit{in} modes $v_k$, with initial conditions at $\etai$, associated with the notion of vacuum of an observer at $\etai$, until $\etaf$. On the other hand, for the \textit{out} modes corresponding to the notion of vacuum of an observer at $\etaf$, it is sufficient to provide its value at that time. 
\subsection{Dynamics of the \textit{in} modes}

At the beginning of inflation, the scale factor grows quasi-exponentially, and the geometry is approximately de Sitter. In a de Sitter universe, the Hubble parameter and curvature scalar are constant, $H=H_0$, $R=12H_0^2$, and the scale factor is given by $a(\eta)=-1/[H_0(\eta-\eta_0)]$. In this scenario, the mode frequency \eqref{eq:ModeFreq} becomes
\begin{equation}
    \omega^2_{k,\dS}=k^2+\frac{\mu^2}{(\eta-\eta_0)^2}, \qquad \mu^2=\frac{m^2}{H_0^2}+12\left(\xi-\frac{1}{6}\right),
\label{eq:FreqDS}
\end{equation}
such that the mode equation \eqref{eq:ModeEq} can be transformed into a Bessel equation of order $\nu=\sqrt{1/4-\mu^2}$, whose solutions are combinations of the Hankel functions $H^{(1)}_{\nu}$ and $H^{(2)}_{\nu}$. In particular, the solution corresponding to the Bunch-Davies vacuum reads \cite{Birrell1982}
\begin{equation}
    v_{k,\text{BD}}(\eta)=\sqrt{\frac{\pi|\eta-\eta_0|}{2}}e^{i\pi \nu/2}H_{\nu}^{(1)}(k|\eta-\eta_0|),
\label{eq:BDModes}
\end{equation}
and behaves as a positive frequency Minkowski mode as $k\abs{\eta} \to \infty$.

We could in principle take Eq. \eqref{eq:BDModes} as initial condition for the modes $v_k$ at $\etai$, from which to solve the mode equation \eqref{eq:ModeEq}. However, this is computationally costly. Instead, we consider the following analytical approximation \cite{Cembranos2023},
\begin{equation}
    v_{k,\text{SR}}=\sqrt{\frac{\pi|\tau_k|}{2}}e^{i\pi n/2} H^{(1)}_{\nu}(k|\tau_k|), \quad \text{with} \quad \tau_k=\frac{\omega_k}{\omega_{k,\dS}}(\eta-\eta_{*,k})+\eta_{*,k}-\eta_0,
\end{equation}
which is valid during slow roll, and compatible with Bunch-Davies initial conditions. Here, $\omega_k$ is given by background dynamics described in Section \ref{sec:backgr}, and $\eta_{*,k}$ marks the limit of validity of the approximation. These mode functions and their derivatives at $\eta_{*,k}$ can then be used as initial conditions to solve \eqref{eq:ModeEq} numerically from $\eta_{*,k}$ to $\etaf$, and obtain the mode functions at the final time, $v_k(\etaf)$. For our choice of $\eta_{*,k}=-500/m_{\phi}$ and $\eta_0=1/H_0$, the error in the initial conditions stays below $1\%$ for the wavenumbers $k$ relevant for pair production in the region of parameter space explored.

\subsection{Choice of \textit{out} vacuum}
\label{subsec:vacch}

To obtain the number density of produced particles, we also need to evaluate the \textit{out} modes $u_k$ at $\etaf$. Recall that the final time is defined as the point from which particle production becomes negligible due to the adiabatic expansion of spacetime. A useful indicator that this is indeed the situation is the condition\footnote{Note that this condition does not completely characterize adiabaticity. Ultimately, one would need to make sure that the number density of produced particles reaches a stationary value \cite{Cembranos2023, Alvarez2025}.}
\begin{equation}
    \left|\frac{\omega'_k(\etaf)}{\omega_k^2(\etaf)}\right|\ll1,
\label{eq:ParamAd}
\end{equation}
which tells us when the time variation of the frequency becomes small. The time at which this condition is fulfilled depends on $k$, $m$, and $\xi$. It happens earlier for increasing mass and decreasing coupling. When condition \eqref{eq:ParamAd} is fulfilled, the adiabatic vacuum prescription provides a reasonable particle notion for computing the density of produced pairs \cite{Birrell1982},
\begin{equation}
    u_k(\etaf)=\frac{1}{\sqrt{\omega_k(\etaf)}}, \qquad u'_k(\etaf)=-\frac{1}{\sqrt{\omega_k(\etaf)}}\left[i\omega_k(\etaf)+\frac{\omega'_k(\etaf)}{2\omega_k(\etaf)}\right].
\label{eq:AdiabaticVacuum}
\end{equation}
With this, we can proceed and evaluate Eq. \eqref{eq:TotalDensity}.

However, when the scalar field mass $m$ is much smaller than the inflaton mass $m_{\phi}$, the time at which condition \eqref{eq:ParamAd} is fulfilled becomes exceedingly large. To avoid such computational cost, we use a different effective vacuum prescription that approximates well the number density of particles computed with the above prescription, even when evaluated at a time $\eta_{\av}\ll\etaf$. This \textit{averaged vacuum} is characterized by the modes with initial conditions
\begin{equation}
    u_{k,\av}(\eta_{\av})=\frac{1}{\sqrt{\omega_{k,\av}(\eta_{\av})}}, \quad 
    u'_{k,\av}(\eta_{\av})=-\left(i\omega_{k,\av}(\eta_{\av})+\frac{\omega'_{k,\av}(\eta_{\av})}{2\omega_{k,\av}(\eta_{\av})}\right)\frac{1}{\sqrt{\omega_{k,\av}(\eta_{\av})}},
    \label{eq:AveragedVacuum}
\end{equation}
where the frequency $\omega_{k,\av}$ is defined by averaging the Ricci scalar oscillations during reheating, using the fact that the Ricci amplitude decays as $1/t^2$ within this regime. This allows us to write 
\begin{equation}
    \omega^2_{k,\av}(\eta)=k^2+a^2(\eta)\left[m^2+\left(\xi-\frac{1}{6}\right)\langle R(\eta)\rangle \right],
\end{equation}
where $\langle R(\eta)\rangle = \mathcal{C}/t^2(\eta)$ denotes the amplitude of the oscillations well inside the reheating epoch, and $\mathcal{C}$ a constant factor. For each potential, $\mathcal{C}$ is calculated by evaluating the Ricci scalar at a late-time $\eta_{\mathcal{C}}$  corresponding to a maximum of the oscillation, so that $\langle R(\eta_{\mathcal{C}})\rangle\equiv R(\eta_{\mathcal{C}})$.

In practice, we solve the mode equation equation \eqref{eq:ModeEq} until a time $\bar{\eta}$ for all pairs $(m, \xi)$ in the considered parameter space. If $\abs{\omega_k^{\prime}/\omega_k^2}\leq 0.01$ at that time for all the wavenumbers $k$ of interest (and therefore condition \eqref{eq:ParamAd} is fulfilled), we use the adiabatic prescription, identifying $\bar{\eta}$ with $\etaf$. Otherwise, we use the averaged vacuum prescription and identify $\bar{\eta}$ with $\eta_{\text{av}}$. The values of $\bar{\eta}$ are $1.6\times 10^{3}m_{\phi}$ for the quadratic scenario and $3.6\times10^2m_{\phi}$ for the Starobinsky potential, leading to $\mathcal{C}=5.4$ and $\mathcal{C}=12.8$, respectively. To ensure that the averaged vacuum provides the right number density of produced particles, we consider the worst case scenario---i.e., the values of $m$ and $\xi$ leading to the least adiabatic frequency---and compare both results. Let us stress that this is nothing but a method for obtaining the number density of produced particles at $\etaf$, which is the physical quantity we are interested in. More details on the validity of this prescription can be found in \cite{Cembranos2023}. 

\section{Spectra and abundance of produced particles}
\label{sec:abundance}

In the following, we present the resulting pair production spectra for the dark matter test field under both quadratic and Starobinsky inflation. Note that we work in terms of physical wavenumbers $k_{\text{phys}}=k/\af$, since the comoving number density in Eq. \eqref{eq:NumberDensity} depends on the scale factor normalization---which we chose differently for each model because of computational reasons.

\begin{figure}[t!]
    \centering
    \includegraphics[scale=0.5]{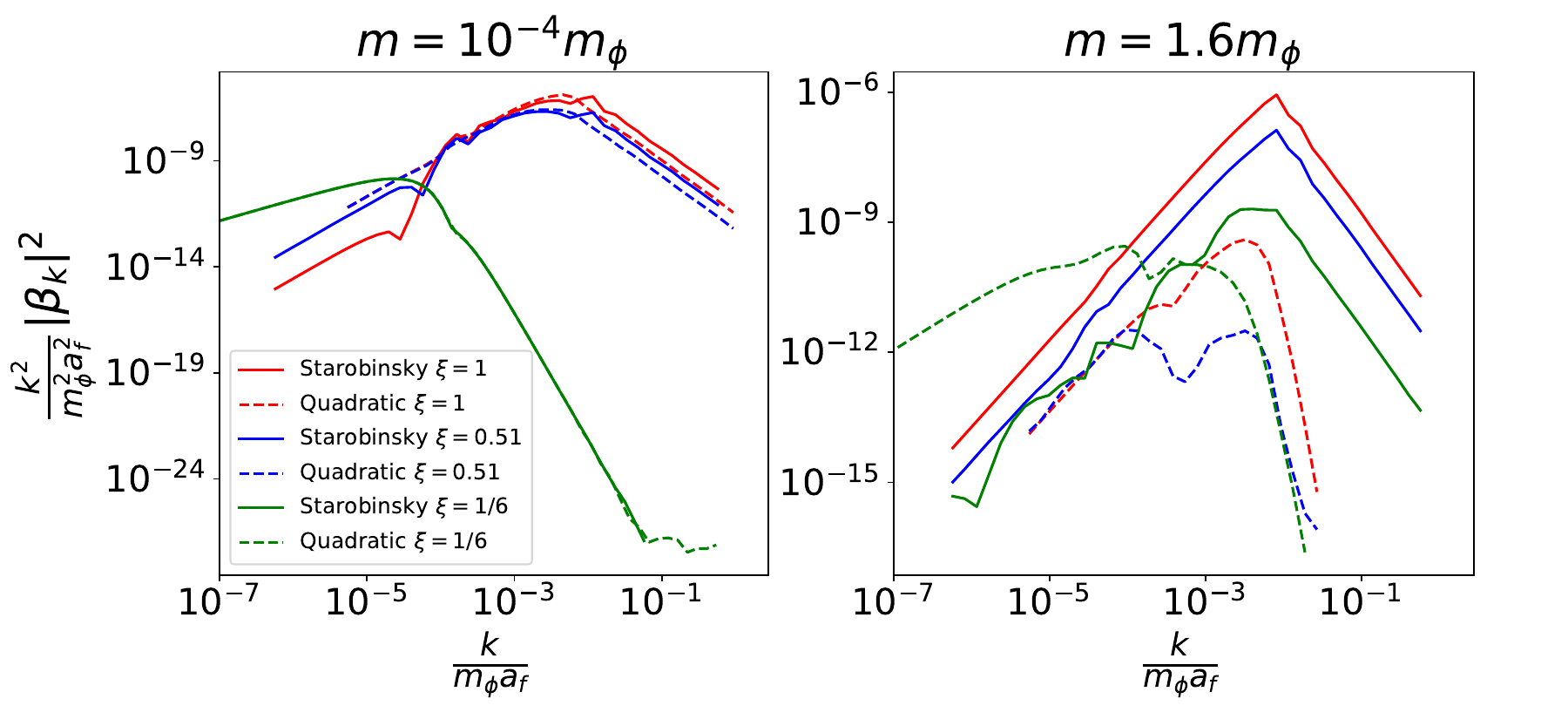}
    \caption{Spectra of produced particles as function of the wavenumber $k$ for $m=10^{-4}m_{\phi}$ (left) and $m=1.6m_{\phi}$ (right), and different values of the coupling $\xi$. The solid lines correspond to the Starobinsky potential, whereas the dashed lines are the results in the quadratic case.}
    \label{fig:SpectraXis}
\end{figure}

First, we study the spectra of produced particles $k^2\abs{\beta_k}^2$, i.e., the integrand of Eq. \eqref{eq:TotalDensity}. In Figure \ref{fig:SpectraXis} we show the corresponding production spectra as function of the wavenumber $k$ for different values of the coupling to the geometry $\xi$, and two very different masses $m$. The solid lines corresponds to the Starobinsky potential, whereas the quadratic results are described by the dashed lines. The left panel shows the results for $m=10^{-4}m_{\phi}$, whereas the right panel corresponds to a much larger $m=1.6m_{\phi}$, above the scale of the quadratic inflaton mass. As a general rule, increasing the value of the coupling also leads to higher amplitudes in the spectra, since the effect of the geometry becomes larger. The behavior is, however, more involved in the region around the inflaton mass, as we can see in the right panel of the figure. Another important point is that the maximum shifts towards higher wavenumbers when increasing $\xi$. The quadratic spectra are compatible with previous results such as those presented in \cite{Cembranos2023}. Notably, for sufficiently low masses, both potentials yield very similar spectra.

\begin{figure}[t!]
    \centering
    \includegraphics[scale=0.5]{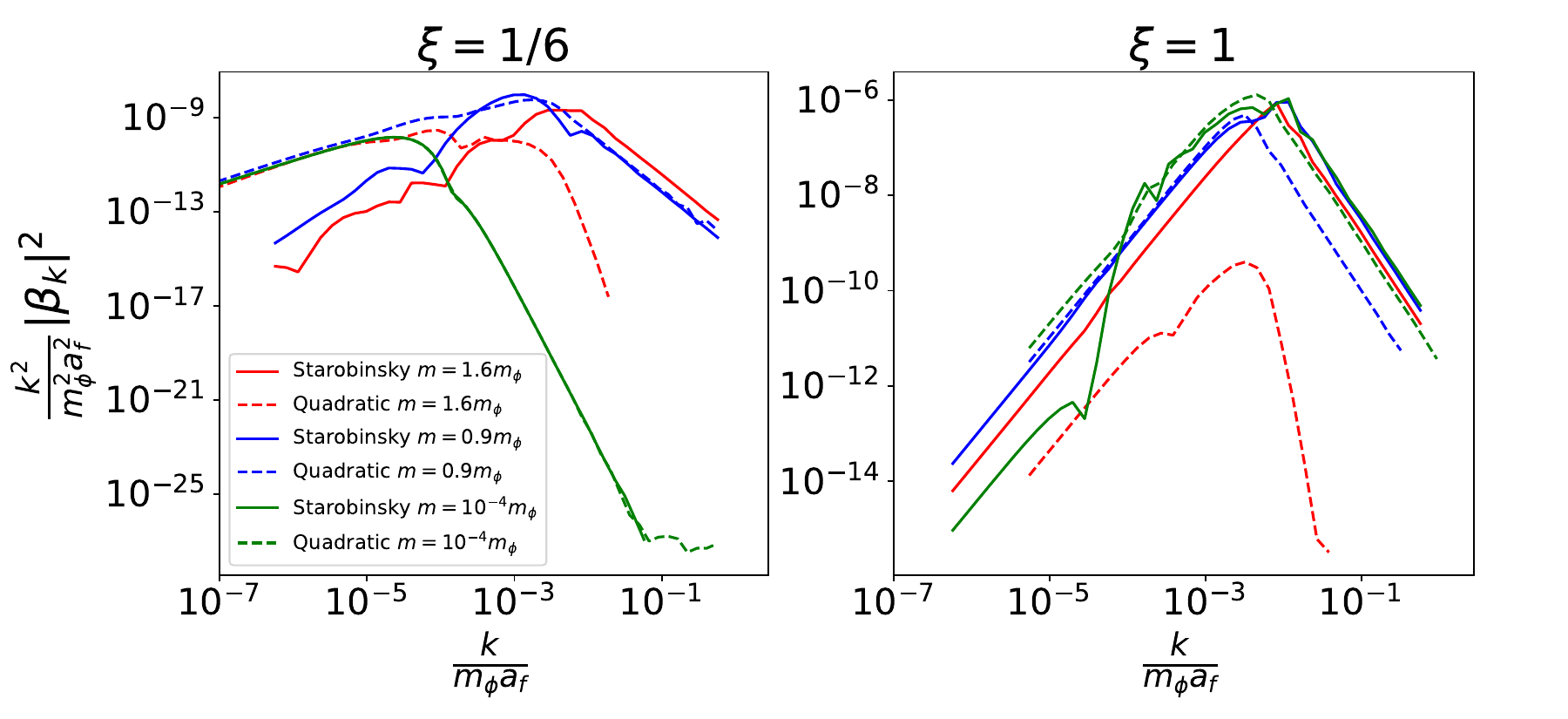}
    \caption{Spectra of produced particles as function of the wavenumber $k$ for $m=10^{-4}m_{\phi}$ (left) and $m=1.6m_{\phi}$ (right), and different values of the coupling $\xi$. The solid lines correspond to the Starobinsky potential, whereas the dashed lines are the results in the quadratic case.}
    \label{fig:SpectraMasses}
\end{figure}

In Figure~\ref{fig:SpectraMasses}, we study how production changes with mass for fixed $\xi=1/6$ (left) and $\xi=1$ (right). We observe that, in general, spectral amplitudes are lower for the conformal case. The differences between the two inflationary potentials appear for $m\gtrsim m_{\phi}, \bar{m}_{\phi}$, close to the scale of inflation.

The total density of produced particles as defined in equation \eqref{eq:TotalDensity} is obtained after integrating the spectra. If we assume that the dark matter field does not interact with any other particles, and gravitational production becomes negligible from $\etaf$ onward, the physical density today can be computed as $n/a_{\td}^3$, taking into account the dilution due to the expansion of spacetime. In practice, we will compute the physical density at the end of reheating, $n/a_{\rm{rh}}^3$, and parametrize the dilution from $\eta_{\text{rh}}$ until today with the reheating temperature $T_{\text{rh}}$. The time at which reheating ends, $\eta_{\text{rh}}$, can be found for a given reheating temperature by taking into account that at this moment radiation dominates, and the Hubble rate can be written as
\begin{equation}
    H^2=\frac{8\pi^3}{90M_P^2}g^{*S}_{\rm{rh}}T_{\rm{rh}}^4=\frac{8\pi}{3M_P^2}\left(\frac{\dot{\phi}(\eta_{\rm{rh}})^2}{2}+V[\phi(\eta_{\rm{rh}})]\right),
\end{equation}
where $g^{*S}$ denotes the number of relativistic degrees of freedom. 

\begin{figure}[t!]
    \centering
    \includegraphics[scale=0.45]{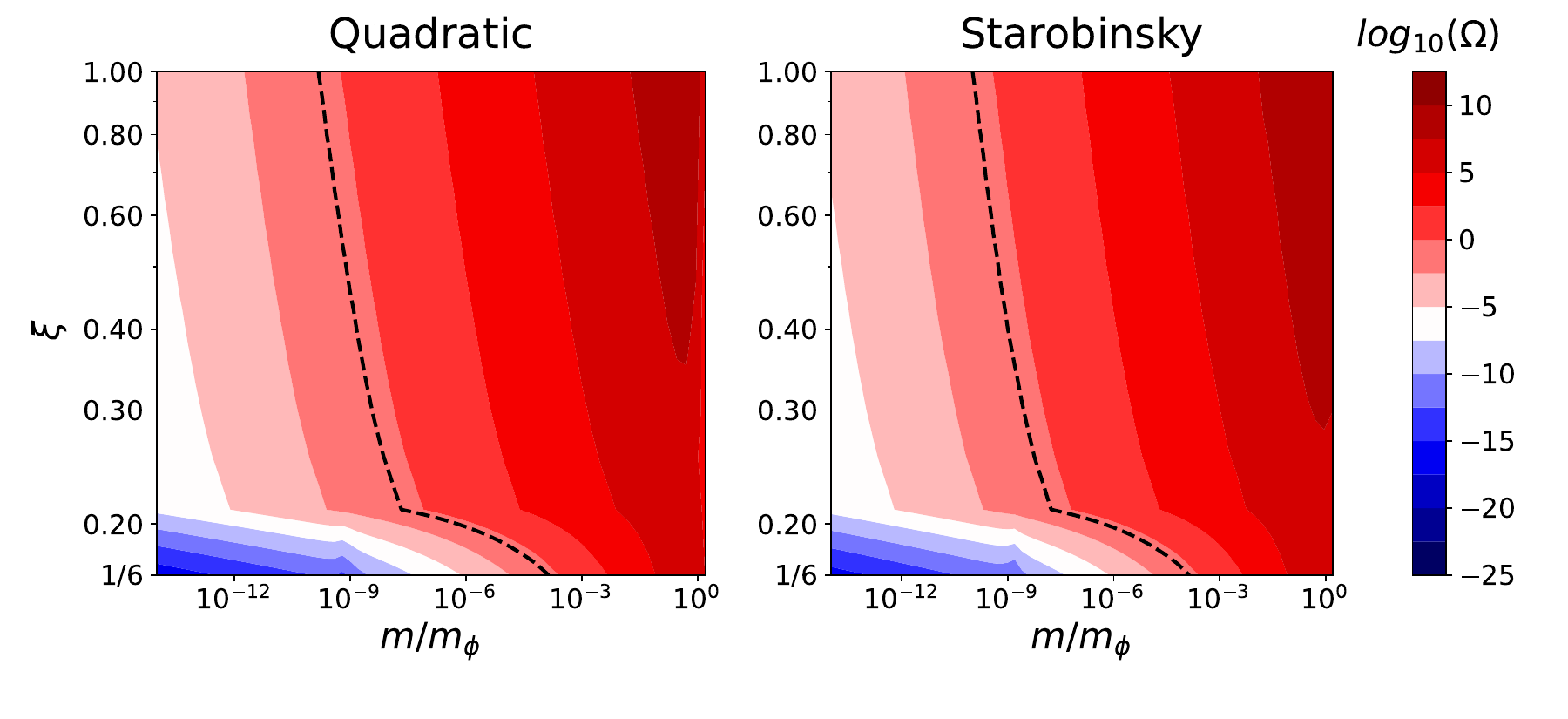}
    \caption{Abundance in the present day of produced particles for each potential as a function of their mass $m$ and coupling $\xi$, for a reheating temperature of $T_{\text{rh}}=10^{13}\ \mathrm{GeV}$. The dashed line corresponds to the observed abundance of dark.}
    \label{fig:AbundancesXiVSm}
\end{figure}

Thus, the corresponding abundance $\Omega$ today can be written as \cite{Cembranos2023}
\begin{equation}
    \Omega(m,\xi)=\frac{8\pi}{3M_P^2H_{\td}^2}\frac{g_{\td}^{*S}}{g_{\rm{rh}}^{*S}}\left(\frac{T_{\td}}{T_{\rm{rh}}}\right)^3m\frac{n(m,\xi)}{a_{\rm{rh}}^3}.
\label{eq:Abundance}
\end{equation}
Note that we have assumed that the dark matter particles are non-relativistic today to calculate the energy density as the product of their number density and mass. 

The time $\etaf$, at which the adiabatic regime is reached for each combination of parameters limits the earliest reheating time $\eta_{\text{rh}}$ we can consider, since we assume that the dynamics are the same until $\etaf$. However, even if reheating ends before adiabaticity is reached, one can argue that the number density of particles produced after reheating is negligible, since spacetime expansion becomes even more adiabatic. Therefore, in practice, we only require $\eta_{\text{}} \leq \eta_{\text{rh}}$, which sets the highest reheating temperature allowed to $T_{\text{rh}} \sim 10^{13} \, \text{GeV}$.  

\begin{figure}[t!]
    \centering
    \includegraphics[scale=0.38]{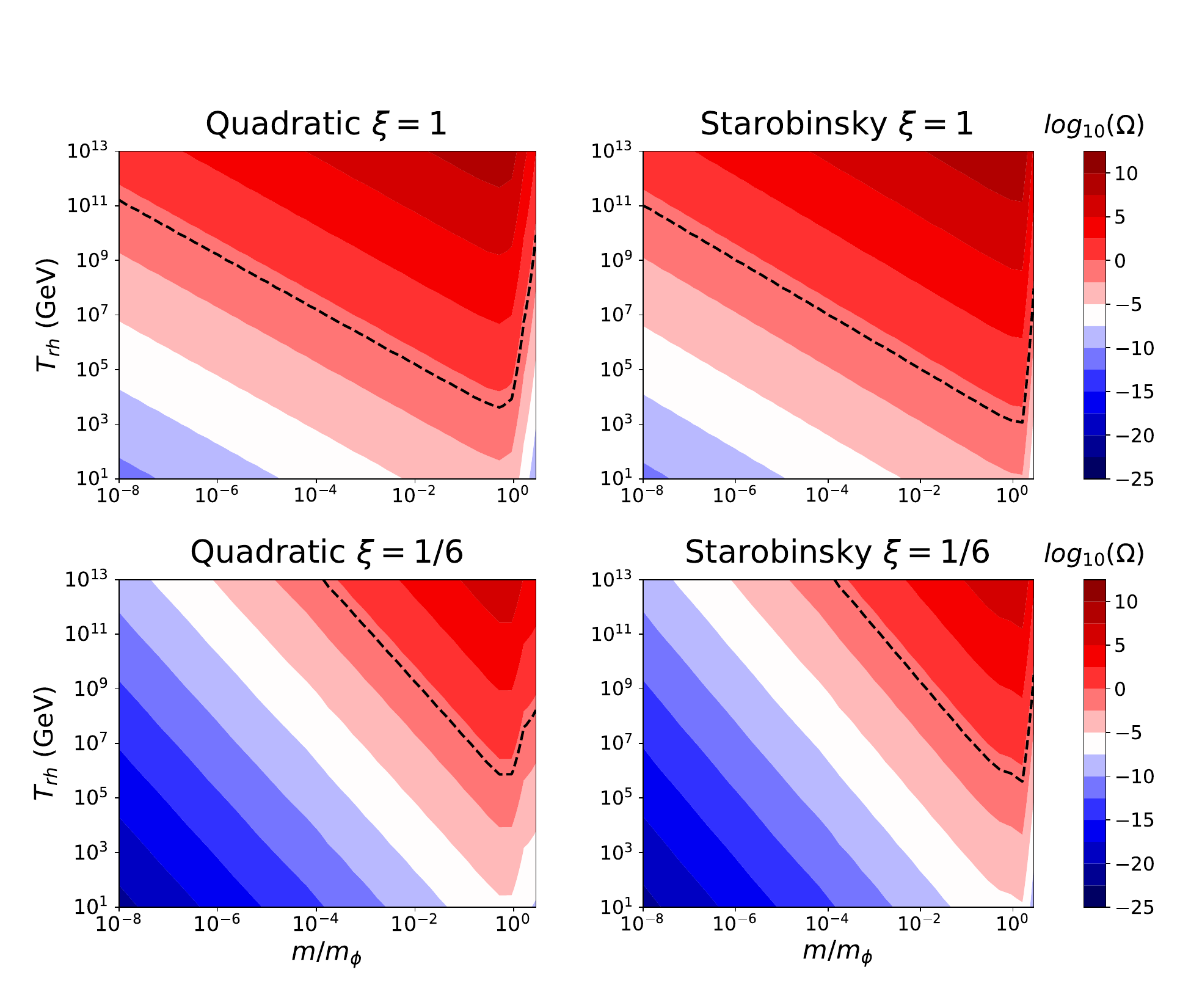}
    \caption{Abundance of produced particles as a function of reheating temperature $T_{\text{rh}}$ and mass $m$, for each potential and two values of the coupling, $\xi=1$ (top) and $\xi=1/6$ (bottom). The dashed line corresponds to the observed abundance of dark matter.}
    \label{fig:AbundancesTVSm}
\end{figure}
\begin{figure}[t!]
    \centering
    \includegraphics[scale=0.38]{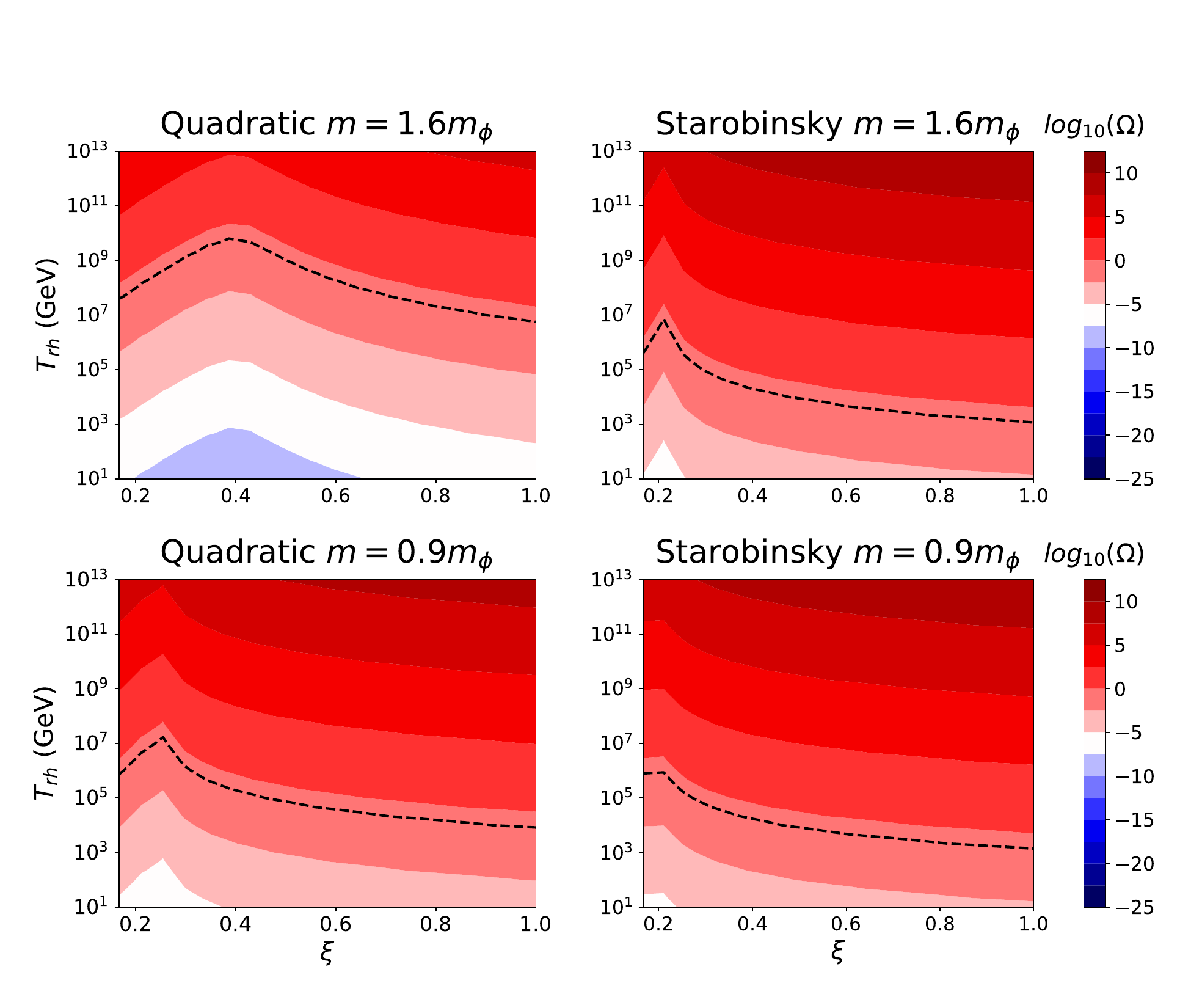}
    \caption{Abundance of produced particles as a function of reheating temperature $T_{\text{rh}}$ and coupling $\xi$, for each potential and two values of mass, $m=1.6m_{\phi}$ (top) and $m=0.9m_{\phi}$ (bottom). The dashed line corresponds to the observed abundance of dark matter.}
    \label{fig:AbundancesTVSxi}
\end{figure}

In Figure \ref{fig:AbundancesXiVSm} we show the abundances in the present day corresponding to the quadratic (left) and Starobinsky (right) models, as function of the dark matter mass $m$ and its coupling to the geometry $\xi$, for a reheating temperature of $T_{\text{rh}}=10^{13} \, \text{GeV}$. Note that we consider abundances larger than the observed amount of dark matter, as other non-standard cosmological evolutions could further dilute the number density of produced particles. Then, the observed amount of dark matter could be recovered in this region of parameter space. Crucially, the resulting abundance is very similar except in the $m \gtrsim m_{\phi}, \bar{m}_{\phi}$ region, when mass of the spectator field is of the order of the inflaton mass. This is expected, as the typical scale of inflation is directly related to this quantity. When the mass of the dark matter candidate becomes of the order of the inflaton mass, production becomes inefficient. Since Starobinsky's inflaton mass is larger, this behavior is expected at higher masses than in the quadratic case.

Note that today's abundance is sensitive to the value of the reheating temperature, since the dilution of the density of dark matter particles is different before and after reheating ends. As Figures \ref{fig:AbundancesTVSm} and \ref{fig:AbundancesTVSxi} show, the abundance increases with increasing temperature, i.e., the number density of produced particles dilutes faster during reheating (when the scale factor behaves as $a(t) \sim t^{2/3}$). For fixed $\xi$, we observe that the abundance increases with mass until $m \sim m_{\phi}, \bar{m}_{\phi}$. After this, increasing the mass of the spectator field results in a lower abundance. In fact, $m$ becomes larger than the characteristic scale of the inflation, $H_0 \simeq 10^{13} \, \text{GeV}$, and parametric resonance becomes inefficient. On the other hand, increasing the coupling $\xi$ to the geometry leads, in general, to more production---except for $m \gtrsim m_{\phi}$, as can be seen in Figure \ref{fig:AbundancesTVSxi}.

\begin{figure}[t!]
    \centering
    \hspace*{-0.75cm}\includegraphics[scale=0.45]{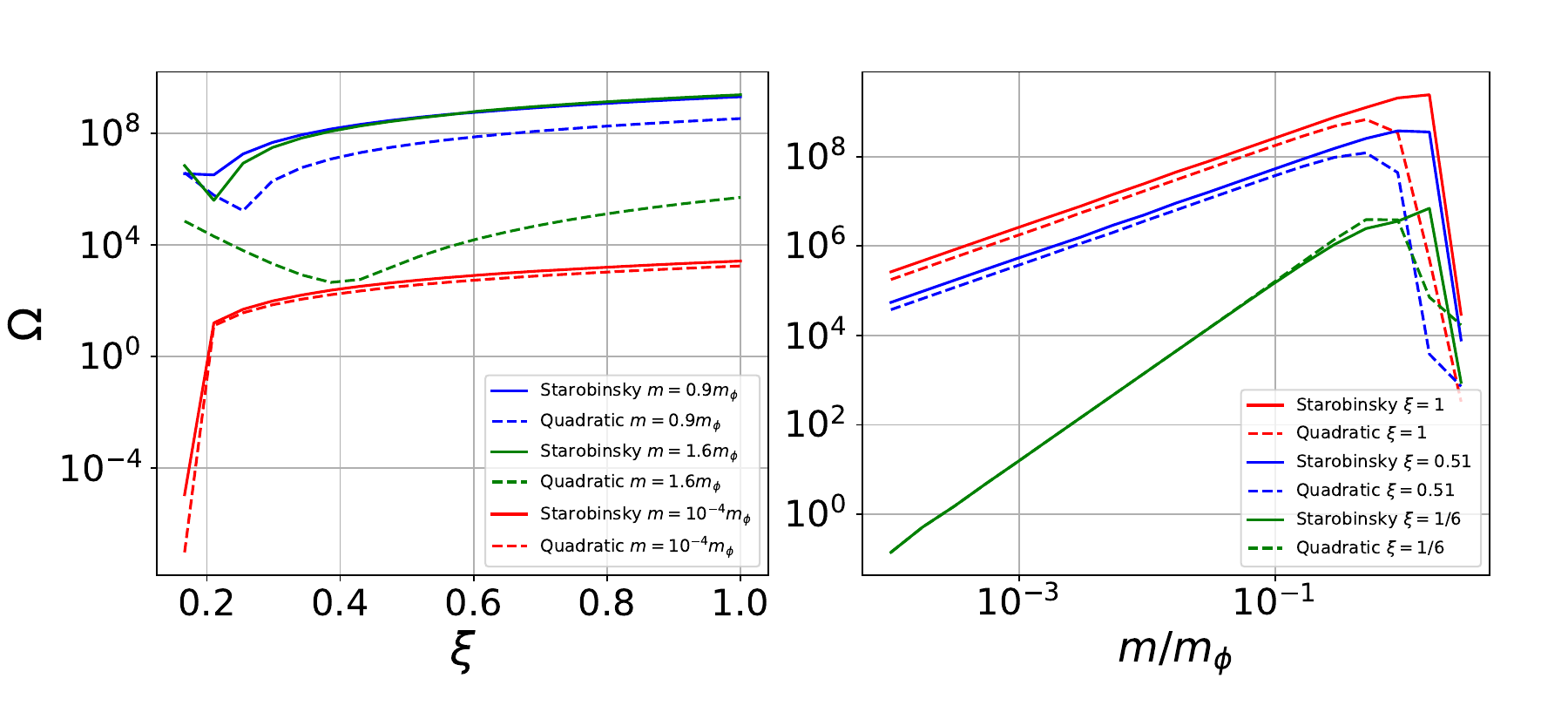}
    \caption{Abundance as a function of coupling $\xi$ (left) and mass $m$ (right), for several values of the other parameter, $T_{rh}=10^{13}\ \mathrm{GeV}$ and each potential. Solid lines correspond to the Starobinsky case, whereas dashed lines denote the results for the quadratic potential.}
    \label{fig:AbundanceVSXim}
\end{figure}

\begin{figure}[t!]
    \centering
    \includegraphics[scale=0.35]{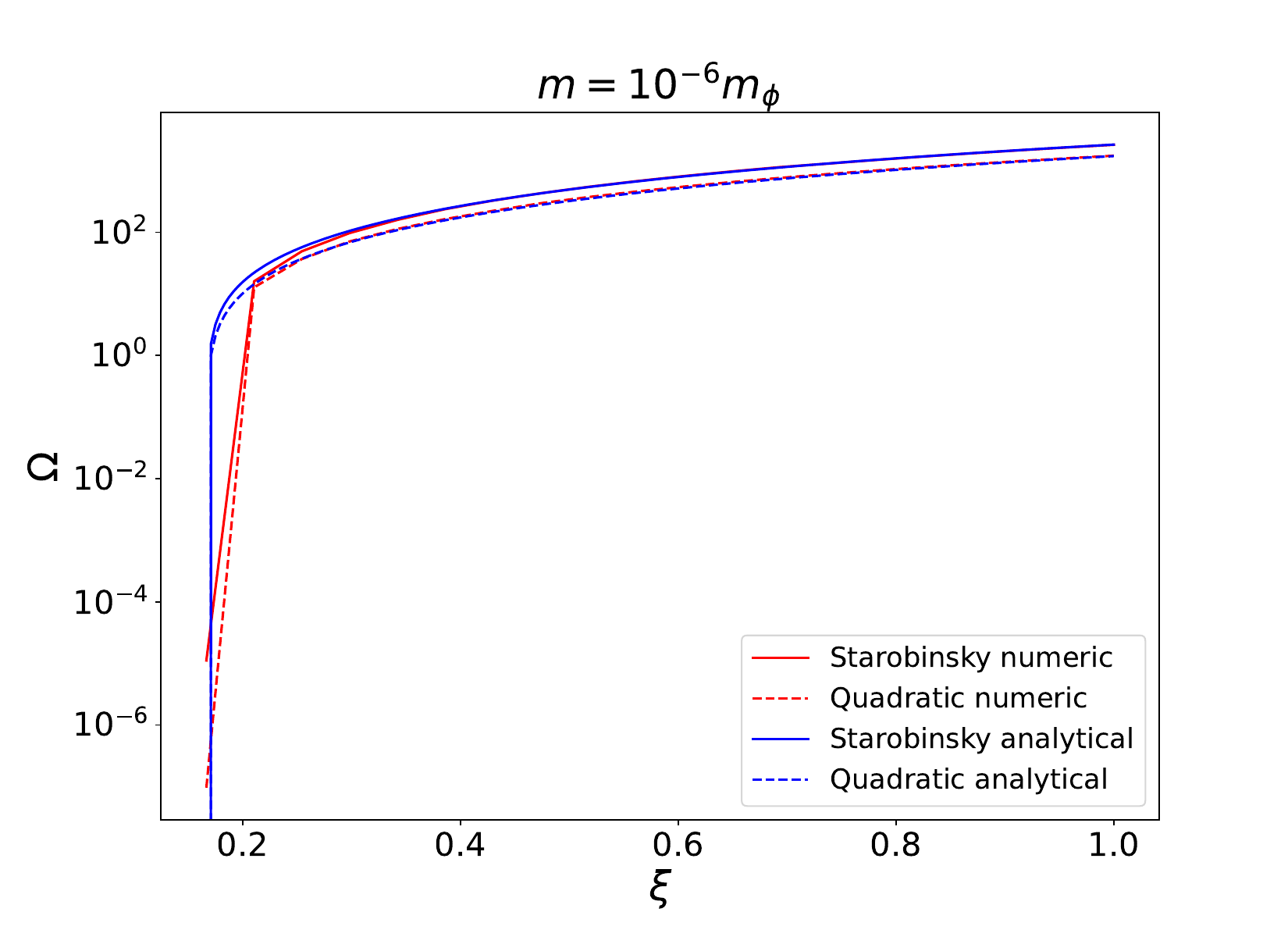}
    \caption{Comparison between our fitting formula (blue) and numerical results (red) of abundance as a function of coupling $\xi$ for a low mass value and $T_{\text{rh}}=10^{13}\ \mathrm{GeV}$. Solid lines correspond to the Starobinsky case, whereas dashed lines denote the results for the quadratic potential.}
    \label{fig:AnalyticApproximation}
\end{figure}

Finally, we compare more closely the two models in Fig. \ref{fig:AbundanceVSXim}, where we show the abundance for several values of dark matter masses as function of the coupling $\xi$ (left), and as function of the mass $m$ for different values of the coupling (right). We observe that abundances in the Starobinsky model are typically larger. There are larger differences when one reaches the region around $m\sim m_{\phi}$ because the production drops in the quadratic case. However, we expect the opposite behavior for $m \sim \bar{m}_{\phi} \simeq 2.3m_{\phi}$, i.e., when reaching the Starobinsky potential mass. The differences also become more pronounced away from the conformal coupling, $\xi=1/6$, since then the curvature scalar terms in the frequency \eqref{eq:ModeFreq} become non-zero. Crucially, both models yield very similar abundances along all the parameter space, as long as one sits sufficiently away from the mass scales that we just discussed.

As expected from the mode frequency equation \eqref{eq:ModeFreq}, for sufficiently small values of the dark matter mass, the number density $n$ only depends on $\xi$, except for small values of the coupling such that $\xi\approx 1/6$. Therefore, in this regime, the mass dependency in the abundance is solely due to the $m$ factor in \eqref{eq:Abundance} coming from the density of non-relativistic particles. We find that, in this regime, the dependence of $n$ on $\xi$ can be well approximated by a quadratic relation. This allows us to write an approximated, analytic expression fitting the abundance as a function of $m$ and $\xi$ for low masses and sufficiently large couplings. Notably, we also find that the abundance for each inflationary potential scales with the square root of its characteristic mass $m_{\text{inf}}$. The corresponding expression for the abundance in the low mass regime reads
\begin{equation}
    \Omega \simeq 2.79\cdot10^{8} \left(\frac{m}{m_{\phi}}\right)\left(\frac{T_{\text{rh}}}{m_{\phi}}\right)\sqrt{\frac{m_{\text{inf}}}{m_{\phi}}}\left(\xi-1/6\right)\left[1+9.60(\xi-1/6)  \right].
\label{eq: analytical fit abundance low m}
\end{equation}
Remember that we are using $m_{\text{inf}}=m_{\phi}=1.21 \times 10^{13} \, \text{GeV}$ for the inflaton mass in the quadratic potential scenario and $m_{\text{inf}}=\bar{m}_{\phi}=2.87 \times 10^{13} \, \text{GeV}$ for the Starobinsky case.

In Figure \ref{fig:AnalyticApproximation}, we show both the numerical abundances in the Starobinsky and the quadratic cases, together with the one obtained using the analytical approximation \eqref{eq: analytical fit abundance low m}, for $m=10^{-6}m_{\phi}$. The differences between this analytic form and our numerical estimations fall below $10\%$ for $\xi-1/6 \gtrsim 0.2$.

\section{Conclusions}
\label{sec:conclusions}

In recent years, gravitational production during inflation and reheating has emerged as a viable and compelling mechanism for generating dark matter. Due to its universality, it has been explored across a range of inflationary contexts, and for different types of dark matter candidates. While spontaneous pair creation in an expanding background occurs for any quantum field coupled to the geometry, the resulting number density produced is sensitive to the details of the inflationary dynamics and reheating history.

In this work, we took a first step toward clarifying the uncertainties associated with the inflationary background in this class of dark matter production models. We focused on two representative single-field inflationary potentials---quadratic and Starobinsky---in the context of a spectator scalar field that is non-minimally coupled to the geometry. We found that the resulting abundances in both models are of the same order of magnitude, as long as the dark matter candidate's mass remains below the corresponding inflaton mass, which sets the characteristic Hubble rate during inflation. This holds true even when the coupling to the curvature is significantly non-conformal---provided it is small enough to neglect backreaction---, though in that case, the sensitivity to the inflationary potential is maximized. In such regimes, the order of magnitude of the resulting abundances appears largely insensitive to the specific inflationary potential considered.

Via the Bogoliubov formalism, we confirmed several results previously highlighted in the literature, including the relevance of the Ricci scalar oscillations and the tachyonic regimes for pair production, and the importance of a consistent choice of \textit{in} and \textit{out} vacua. Furthermore, we showed that the uncertainty in the value of the reheating temperature $T_{\text{reh}}$ allows for a wide range of parameter space compatible with the experimentally observed dark matter abundance. Notably, we derived an analytic expression for today's dark matter abundance as predicted by the gravitational particle production mechanism, as a function of the candidate's mass $m$, the non-minimal coupling $\xi$, and the inflaton mass of the corresponding potential under study $m_{\text{inf}}$. This approximation provides an excellent fit to our numerical results, with deviations below $10 \%$, for the regime of low masses ($m < m_{\text{inf}}$), and couplings above the conformal point ($\xi-1/6 > 0.2$).

In the future, we aim to investigate other types of spectator fields to test how particle production depends on the underlying inflationary dynamics. While some modes exhibit more complex behavior---such as the longitudinal mode of a vector field \cite{Cembranos2024}---, others, like the transverse modes of vector fields, behave analogously to scalar fields. In such cases, we expect our conclusions to carry over without significant modification. A more comprehensive analysis would require to also examine qualitatively different inflationary potentials, especially those that do not behave quadratically during reheating. Although definitive statements require explicit calculations, we expect that for conformally coupled fields, pair production proceeds in a similar fashion, as curvature oscillations play no significant role in this case.

In summary, we have provided a quantitative estimate of gravitational production uncertainties arising from the specifics of the inflationary and reheating stages. This constitutes a first step toward a more comprehensive analysis of the model dependence of such dark matter production mechanism, which we aim at realizing in subsequent works---also including more exhaustive constraints on other cosmological observables, coming from, e.g., isocurvature or structure formation.

\section*{Acknowledgements}

This work is partially supported by the project PID2022-139841NB-I00 funded by MICIU/AEI/10.13039/50110001 1033 and by ERDF/EU, and by the R+D+I Project PID2023-149018NB-C44, funded by MICIU/AEI/10.13039/501100011033 and by ERDF/EU. This work is also part of the COST (European Cooperation in Science and Technology) Actions CA21106, CA21136, CA22113 and CA23130. Additionally, A.P.L. also acknowledges support through MICIU fellowship FPU20/05603. J.O.R. acknowledges financial support from the FPI fellowship PREP2023-002243 of the Spanish Ministry of Science, Innovation and Universities. He  is also  supported by the Spanish grant PID2023-149016NBI00 (funded by MCIN/AEI/10.13039/501100011033 and by “ERDF A way of making Europe”) and the Basque government Grant No. IT1628-22 (Spain).


\bibliographystyle{JHEP.bst}
\bibliography{article2.bib}

\end{document}